\documentclass[twocolumn,showpacs,preprintnumbers,amsmath,amssymb]{revtex4}
\usepackage{graphicx}
\usepackage{dcolumn}
\usepackage{bm}

\begin{document}

\title{Direct Measurement of intermediate-range Casimir-Polder potentials}

\author{H. Bender}
\author{Ph. W. Courteille}
\author{C. Marzok}
\author{C. Zimmermann}
\author{S. Slama}
\affiliation{Physikalisches Institut, Eberhard Karls Universit\"at T\"ubingen,\\
Auf der Morgenstelle 14, D-72076 T\"ubingen, Germany}

\date{\today}

\begin{abstract}
We present the first direct measurements of Casimir-Polder forces
between solid surfaces and atomic gases in the transition regime
between the electrostatic short-distance and the retarded
long-distance limit. The experimental method is based on ultracold
ground-state Rb atoms that are reflected from evanescent wave
barriers at the surface of a dielectric glass prism. Our novel
approach does not require assumptions about the potential shape. The
experimental data confirm the theoretical prediction in the
transition regime.

\end{abstract}


\maketitle Although being the most precisely tested theory in
physics, quantum electrodynamics (QED) leads only in very special
cases to measurable forces. One example are the well-known Casimir
and van der Waals forces \cite{Casimir48, Lifshitz61}. In addition
to their fundamental importance, a detailled understanding of these
forces is crucial for testing new fundamental physics at short
distances such as non-Newtonian gravitational forces
\cite{Onofrio06, Decca07}. Furthermore, they have important
technological implications for the development of micromachines with
nanoscale moving parts \cite{Buks01, DelRio05}. Today, Casimir
forces between solids can be measured with high precision
\cite{Chen06, Hertlein08, Corwin09}. These measurements are all done
with objects that are large compared to the relevant distances where
Casimir forces become dominant. Therefore, the underlying theory
contains the macroscopic properties of the objects, i.e. the
dielectric functions. Moreover, also the geometry of the macroscopic
bodies plays an important role due to the non-additivity of Casimir
forces. A much cleaner situation is given when the test object is
microscopic. This is in good approximation true for a single atom.
In this case the force by which the atom is attracted towards a
surface is often referred to as Casimir-Polder (CP) force. The shape
of the CP potential depends on the distance from the surface. While
in the limits of short and long distances CP-forces can be
approximated by different power laws \cite{LennardJones32,
Casimir48}, in the transition regime simple analytical expressions
do not exist. Here, the potential is given by an integral over the
frequency-dependent dielectric function of the substrate and the
polarizability of the atom \cite{Scheel08}.\\

  \begin{figure}[ht]
        \centerline{\scalebox{0.8}{\includegraphics{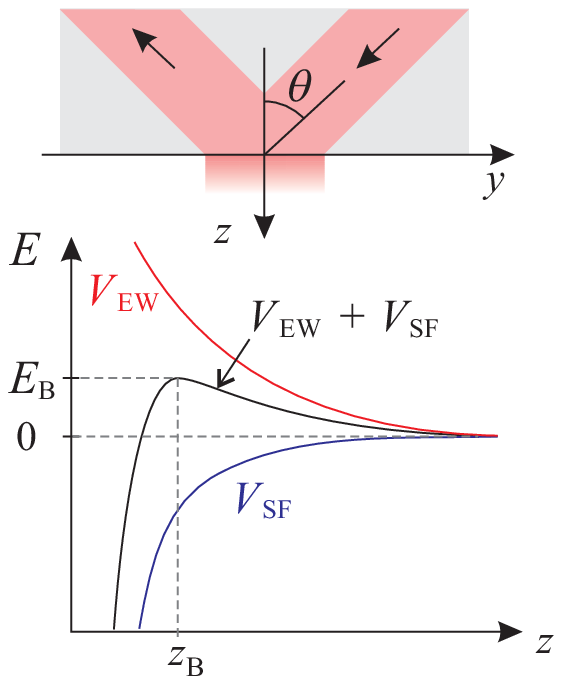}}}\caption{
        Experimental situation. A repulsive evanescent wave potential $V_{\textrm{EW}}$ and an
        attractive surface potential $V_{\textrm{SF}}$ sum up to build a barrier
        at a distance $z_B$ from the surface with height $E_B$.}
        \label{fig:setup}
    \end{figure}
In the last two decades many efforts have been made to measure CP
forces including sophisticated approaches such as diffraction and
interferometry of cold atom beams at thin transmission gratings
\cite{Toennies99, Cronin05} and quantum reflection of atoms from
solid surfaces
\cite{Shimizu01,Druzhinina03,Pasquini04,Schoellkopf08}. In these
experiments CP forces were studied indirectly by fitting the
coefficients of the theoretical surface potentials to the measured
data. Direct methods that have been developed up to now can be
divided into spectroscopic \cite{Hinds92, Hinds93, Failache99,
Fichet07} and kinetic measurements. The latter include reflection of
cold atoms from an atomic mirror \cite{Aspect96} and surface-induced
frequency-shifts of oscillating Bose-Einstein condensates in
magnetic traps \cite{Obrecht07}. In this article the transition
regime is probed for the first time by a direct model-free
measurement. This is done by reflecting ultracold atoms from
evanescent wave barriers similar to previous work \cite{Aspect96}.
However, here we systematically vary the mirror potential and
introduce a new data analysis which allows for the direct
investigation of surface potentials at sub-wavelength distances from
the surface. The new method makes use of a known repulsive potential
$V_{\textrm{EW}}(z)$ that is added to the unknown attractive surface
potential $V_{\textrm{SF}}(z)$ with the goal to generate a potential
barrier. By varying the strength of $V_{\textrm{EW}}(z)$, the height
and the position of the barrier can be adjusted. The height is
measured by reflecting cold atoms of a given energy. The position
can be determined from the derivative of the barrier height with
respect to the strength of $V_{\textrm{EW}}(z)$. This last step is
the key feature for reconstructing the unknown surface potential.\\

The experimental situation is shown in Fig. \ref{fig:setup}. An
evanescent wave leaking out from the surface of a transparent
substrate generates a repulsive dipole potential of the form
\begin{equation}\label{eq:EWPotential}
V_{\textrm{EW}}=C_0\cdot P\cdot
\exp{\left\{-2\frac{z}{z_0}\right\}}~,
\end{equation}
with a constant $C_0$ and the field decay length $z_0$
\cite{Grimm00}. The total potential is now the sum of the
(attractive) unknown surface potential $V_{\textrm{SF}}$ and the EW
Potential
\begin{equation}
V_{\textrm{tot}}=V_{\textrm{SF}}+V_{\textrm{EW}}~.
\end{equation}
 If the repulsive potential is strong enough, a potential barrier is
formed at a distance $z_B(P)$ from the surface. At the maximum
$V_{\textrm{tot}}^\prime=0$, which means that
\begin{equation}\label{VCPprime}
V_{\textrm{SF}}^\prime(z=z_B)=2C_0\cdot \frac{P}{z_0}\cdot
\exp{\left\{-2\frac{z_B}{z_0}\right\}}~.
\end{equation}
Furthermore, the height of the barrier is given by
\begin{equation}\label{barrierheight}
E_B=V_{\textrm{SF}}(z_B)+C_0\cdot P\cdot
\exp{\left\{-2\frac{z_B}{z_0}\right\}}~.
\end{equation}
In the experiment this barrier height $E_B$ is measured as a
function of the laser power $P$. From the data the derivative
$\frac{dE_B}{dP}$ is taken and compared to theory as follows. We
differantiate (\ref{barrierheight}) with respect to the laser power
taking the inner derivatives into account and substitute eq.
(\ref{VCPprime}) into the result. Note that $z_B(P)$ is a function
of $P$. This delivers the following equation:
\begin{equation}\label{calc2}
\frac{dE_B}{dP}= C_0\cdot\exp{\left\{-2\frac{z_B}{z_0}\right\}~.}
\end{equation}
The crucial point here is that this derivative depends only on the
evanescent wave potential and not on the surface potential. From
this expression the position of the barrier can be calculated to be
\begin{equation}\label{eq:zB(P)}
z_B(P)=-\frac{z_0}{2}\cdot\ln\left(\frac{1}{C_0}\cdot\frac{dE_B}{dP}\right)~.
\end{equation}
With this knowledge of the barrier height $E_B$ and the barrier
position $z_B$ the unknown surface potential $V_{\textrm{SF}}(z_B)$
can be determined by solving (\ref{barrierheight}).\\

\begin{figure}[th]
        \centerline{\scalebox{1.2}{\includegraphics{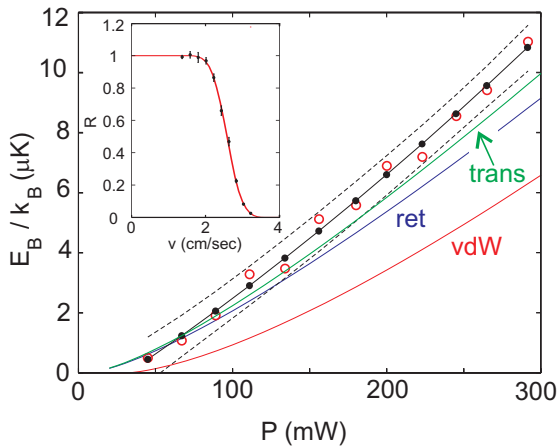}}}\caption{
        Inset: Typical measurement of reflectivity $R$ for a laser power of $P=134$~mW.
        The fitted barrier height is $E_B=k_{\textrm{B}}\times(3.62\pm 0.03)~\mu$K.
        Main figure: Dependence of the
        barrier height on the laser power. The open circles are data points obtained from curves
        similar to the ones shown in the inset. The solid line close to the data points
        is a constraint fit to the data as explained in the text
        with the dashed lines limiting the $95$\% confidence interval.
        The three curves (vdW, ret, trans) are theoretical expectations for the
        barrier height assuming a nonretarded van der Waals-like potential, a
        retarded potential and the full QED potential.}
        \label{fig:BH(P)}
    \end{figure}

The experimental task is to measure the height of the potential
barrier as a function of the laser power. For a given laser power
this is done by classical reflection of an ultracold atom cloud with
variable kinetic energy from the barrier. The experiment is carried
out with a setup explained in detail in \cite{Bender09}. Here only a
short summary is given. An ultracold atomic cloud is prepared in a
Joffe-Pritchard type trap some hundred micrometers below the
superpolished surface of a dielectric glass prism which is mounted
upside down in a vacuum chamber. Almost pure condensates can be
generated with some $10^5$ atoms. For this experiment however, only
very cold thermal clouds are prepared in order to avoid effects due
to the interaction between the atoms at high density.  The ultracold
$^{87}$Rb cloud is held in the magnetic trap at a fixed distance
$z_1$ below the prism surface. Then a vertical laser beam
(wavelength $\lambda=830$~nm) that propagates perpendicularly
through the prism from the top is switched on adiabatically in order
to generate a dipole trap with radial and axial trapping frequencies
of $\omega_{\textrm{d,r}}=2\pi\times50~$Hz and
$\omega_{\textrm{d,z}}<2\pi\times0.1~$Hz. In this combined magnetic
/ dipole trap a cloud temperature of $T\sim100~$nK is measured. Now
the atoms are accelerated towards the surface by a sudden shift of
the magnetic trapping minimum to a new variable position $z_2$.
After a waiting time of a quarter of an oscillation period during
which the atoms accelerate in the shifted trap, the magnetic field
is quickly ramped to a constant gradient which compensates for the
gravitational force. The atoms now move towards the surface with a
nearly constant velocity $v_0$. It is determined by absorption
imaging of the position of the cloud in the first few milliseconds
of its motion. While the atoms move to the surface they are slightly
accelerated due to residual curvature of the levitation potential.
This effect is taken into account as a correction of the measured
velocity. The velocity at the surface is then given by
$v=\sqrt{v_0^2+\omega_{\textrm{lev}}^2z_2^2}$ with
$\omega_{\textrm{lev}}=2\pi\times 4$~Hz. During the reflection of
the atoms the dipole trap guarantees radial confinement. The
measured radial FWHM width of the atomic cloud is $40~\mu\textrm{m}$
such that the atoms are reflected only from the center of the
evanescent wave. There, the potential barrier reaches its maximum
due to the Gaussian intensity distribution of the EW laserspot. The
EW laser is centered around $\lambda=765$~nm with a spectral width
of $\Delta\lambda=\pm1$~nm. The vertical potential generated by the
dipole trap is very weak and can be neglected. After contact with
the surface the number of reflected atoms is determined by
absorption imaging. The measurement is repeated for various trap
displacements $\Delta z=z_2-z_1$ which correspond to different
velocities $v$. Typical results are shown in the inset of Fig.
\ref{fig:BH(P)}. The data points show a gradual decrease from the
situation where all atoms are reflected ($R=1$) to full transmission
$R=0$. The width of this decrease is mainly dominated by the
Gaussian velocity distribution of the atoms corresponding to the
temperature of the cloud. It is also slightly effected by the
imhomogeneous barrier height due to the Gaussian transversal
intensity profile of the EW laser beam. Nonclassical broadening
effects like quantum reflection and tunneling may play a role for
the higher laser powers used, where the surface potential is
steepened by the evanescent wave \cite{Kallush05}. However, its
influence on the center of the decrease and by that on the measured
barrier height is negligible. Thus the data points in
Fig.~\ref{fig:BH(P)} are fit with a model which implements only the
two classical broadening mechanisms mentioned above.
The result of the fits for various laser powers is plotted in the
main part of Fig.~\ref{fig:BH(P)} (open circles). For comparison,
barrier heights are plotted as derived from the theoretical CP
potentials valid in the different regimes:
\begin{equation}\label{eq:VCP1}
V_{\textrm{CP}}=\left\{
{V_{\textrm{vdW}}=-\frac{C_3}{z^3}\hspace{0.5cm}(z\gg l),} \atop
{V_{\textrm{ret}}=-\frac{C_4}{z^4}\hspace{0.5cm}(z\ll l),} \right.
\end{equation}
with a typical distance $l$ separating both regimes. For calculating
the potential coefficients $C_j$ the values for an ideally
conducting surface $C_j^{\textrm{ic}}$ are taken from
\cite{Friedrich02}and corrected for the dielectric surface. In the
case of the $C_4$ coefficient the correction is given by a factor
$\Phi(n)$ with refractive index $n=1.512$ \cite{Lifshitz61}. In the
case of the $C_3$ coefficient an often made approximation leads to a
correction factor of $\frac{n-1}{n+1}$. Our corrected coefficients
are $C_3=5.8\cdot10^{-49}~\textrm{Jm}^3$ and
$C_4=5.4\cdot10^{-56}~\textrm{Jm}^4$. The transition length can be
estimated from the intersection between the retarded and the
nonretarded curve to be $l\approx\frac{C_4}{C_3}=92$~nm. In this
range we calculate the Casimir-Polder potential correctly by solving
the full QED formula (5.39) in \cite{Scheel08}. For this purpose the
magnetic permeability of glass is $\mu(\omega)\equiv\mu=1$ and the
atomic polarizability $\alpha(\omega)$ follows from
\cite{Safronova04}. The dielectric function $\epsilon(\omega)$ is
determined from optical data which are available for glass in a wide
range \cite{Palik85}. The barrier height which is derived from this
correct CP potential fits best to the experimental data, although
also here a deviation is observed. This deviation is particularly
large for high laser powers, where the data exceed the theoretical
values. A model-free comparison between experiment and theory is
possible, if the surface potential is extracted from the measurement
as explained above. The required derivative
$\frac{dE_B}{dP}\approx\frac{E_B(n+1)-E_B(n)}{P(n+1)-P(n)}$ is taken
from a smoothened curve that can be obtained by fitting the measured
barrier heights. The fit function must be chosen very carefully in
order to stay model-free. A polynomial fit function e.g. would
implicitly assume a surface potential of a certain shape. To
maintain generality the values of the fit function at the measured
laser powers are the parameters of the fit. Furthermore, the fit
function fulfils the following three constraints: (1) the first
derivative is positive at each point, (2) the second derivative is
also positive at each point, and (3) the third derivative is
negative at each point. The physical reason for (1) is that an
increasing laser power leads to a growing barrier height. Constraint
(2) is equivalent to the fact that for increasing laser power the
barrier gets closer to the surface and constraint (3) means that the
rate at which the barrier gets closer to the surface decreases.
These assumptions are valid for any attractive surface potential
whose attraction is growing with decreasing distance from the
surface. This is the only assumption on the potential shape we make.
The result of the fit is plotted as dots in Fig. \ref{fig:BH(P)}
\cite{footnote1}. To guide the eye the fit points are linearly
interpolated. From the fit points the surface potential is
determined by equations (\ref{barrierheight}) and (\ref{eq:zB(P)}).
The only parameters in this calculation are the ones describing the
evanescent wave: the field decay length $z_0$ is calculated from a
measured incidence angle of $\theta=43.4^\circ\pm 0.1^\circ$, and a
laser wavelength of $\lambda=765$~nm to $z_0=(430\pm10)$~nm. The
proportionality factor $C_0$ is calculated from standard dipole trap
theory \cite{Grimm00} with a measured beam waist of
$w_{0,\textrm{x}}=(170\pm5)\mu\textrm{m}$ and
$w_{0,\textrm{y}}=(227\pm5)\mu\textrm{m}$. The evanescent wave
intensity is given (for p-polarized light) by
$\left|\frac{E_{\textrm{EW}}}{E_{\textrm{in}}}\right|^2=\frac{1}{n}\cdot\frac{4n^2\cos(\theta)^2}{(\cos(\theta)^2+n^2(n^2\sin(\theta)^2-1))}$,
with the refractive index  $n$ of the prism. With these parameters
the proportionality constant is $C_0=(1.35\pm 0.05)\cdot
10^{-27}~\textrm{J/W}$.
  \begin{figure}[ht]
        \centerline{\scalebox{.75}{\includegraphics{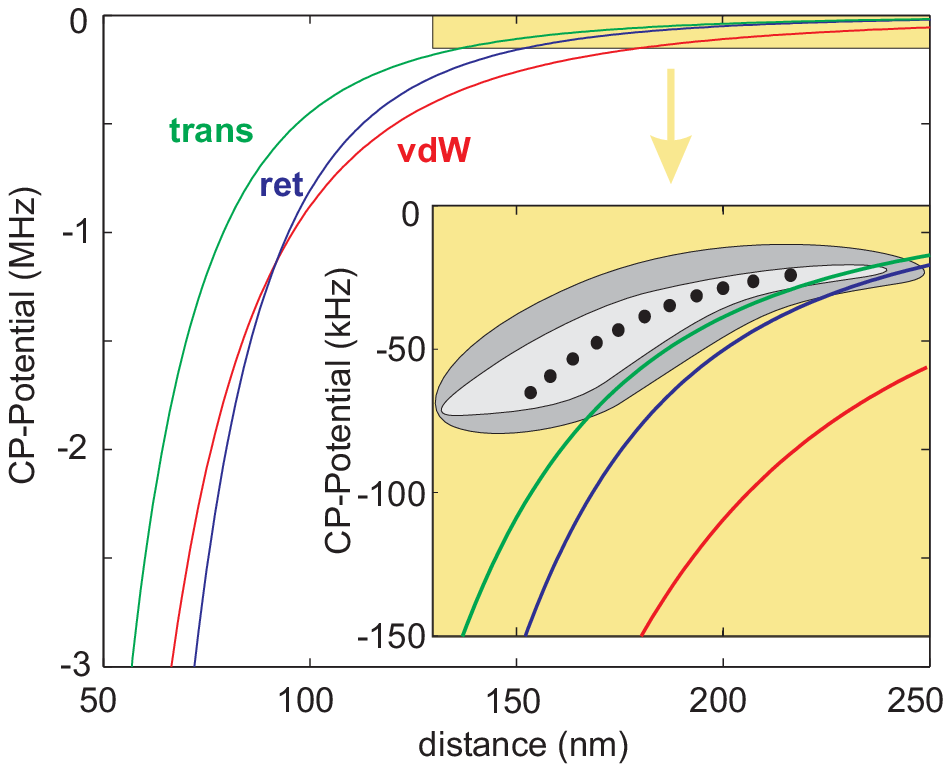}}}\caption{
Measured and theoretical Casimir-Polder potentials: in the large
figure the theoretical surface potentials are plotted, i.e. the
nonretarded van der Waals potential (vdW) and the retarded
Casimir-Polder potential (ret). The full theoretical curve also
valid in the transition regime (trans) approaches the retarded curve
for large distances and the nonretarded curve for small distances.
The inset magnifies the colored box, in which the measured data
points are lying. Statistical and systematic errors are indicated by
the bright and the dark grey shaded area respectively.}
\label{fig:SF-Pot}
\end{figure}

Figure \ref{fig:SF-Pot} shows the surface potential determined from
the measured barrier height (dots). Statistical errors are due to
the spread of the data points in Fig. \ref{fig:BH(P)} around the fit
curve. For the vertical axis they are given by a shift of the fit
curve to the $95$\% confidence interval boundaries. For the
horizontal axis the evaluation of the statistical error depends on
the error of the gradient of the fit curve. An estimation of this
error can be given by calculating the mean gradient and assuming a
maximum gradient by a linear interpolation between the lower left
border of the confidence interval with the upper right border. For
the minimum gradient the upper left border is connected with the
lower right border. This gives an uncertainty in the gradient
$\delta\left(\frac{dE_B}{dP}\right)$. The uncertainty of the barrier
position $\delta z_\textrm{B}$ is then given by means of
(\ref{eq:zB(P)}). Systematic errors are due to the uncertainty for
$z_0$, $C_0$, $P$ and $v$. They turn out to be smaller than the
systematic errors. Taking into account these uncertainties the
measurements agree best with the full QED calculation. The retarded
and the static potential can be excluded. Other experiments show
that patch potentials from adsorbed Rubidium atoms can play a role
in the surface potential \cite{McGuirk04}. Such potentials increase
the attraction between surface and atom. In contrast, our
measurements show a slightly smaller attraction than theoretically
expected. Therefore, patch potentials seem to be negligible in this
setup. This might be due to a permanent exposure of the prism
surface to the evanescent wave which can either lead to laser
induced desorption of atoms \cite{Wilde06} or to an increased
diffusion of atoms on the surface.

In this article direct measurements of the Casimir-Polder force
between ground-state Rb atoms and the surface of a dielectric glass
prism at distances between $160$~nm and $230$~nm are presented. A
novel method has been introduced which is based on a test potential
generated with an optical evanescent wave at the glass surface. The
measurements do not coincide with the limiting formulas valid in the
static and in the retarded regime. However, they do agree with a
correct QED calculation and thus confirm the present theory. In
addition to the mentioned measurement errors the observed small
deviation might be caused by the incomplete knowledge of the
dielectric function of the used borosilikate glass prism. For
calculating the theoretical curve the well-known dielectric function
of SiO$_2$ glass has been used. However, the optical properties of
glasses vary depending on the exact type of glass \cite{Efimov06}.
It is therefore possible that the theoretical curve slightly
deviates from the real situation in the experiment. Already a
moderate increase in the experimental resolution will make it
possible to discern between such theoretical and experimental errors
and might reveal new physics that could be hidden in the observed
deviation.

We acknowledge financial support by the DFG within the EuroCors
program of the ESF. We would like to thank Stefan Scheel and Andreas
G\"unther for helpful discussions.


\begin{thebibliography}{10}
\bibitem{Casimir48}  H.B.G. Casimir and D. Polder, \textit{Phys. Rev.} \textbf{73}, 360 (1948)
\bibitem{Lifshitz61}  I. Dzyaloshinskii, E. Lifshitz, and L. Pitaevskii, \textit{Adv. Phys.} \textbf{10}, 165 (1961).
\bibitem{Onofrio06}  R. Onofrio, \textit{NJP} \textbf{8}, 237 (2006).
\bibitem{Decca07}  R.S. Decca et al., \textit{Phys. Rev. D} \textbf{75}, 077101 (2007).
\bibitem{Buks01}  E. Buks, and M.L. Roukes, \textit{Phys. Rev. B} \textbf{63}, 033402 (2001).
\bibitem{DelRio05}  F.W. DelRio et al., \textit{Nature Mat.} \textbf{4}, 629 (2005).
\bibitem{Chen06}  F. Chen, and U. Mohideen, \textit{J. Phys. A} \textbf{39}, 6233 (2006).
\bibitem{Hertlein08}  C. Hertlein, L. Helden, A. Gambassi, S. Dietrich, and C. Bechinger, \textit{Nature} \textbf{451}, 172 (2008).
\bibitem{Corwin09}  A.D. Corwin, and M.P. deBoer, \textit{JMEMS} \textbf{18}(2), 250 (2009).
\bibitem{LennardJones32}  J.E. Lennard Jones, \textit{Trans. Far. Soc.} \textbf{28}, 333 (1932)
\bibitem{Scheel08} S. Scheel and S.Y. Buhmann, \textit{acta phys. slov.} \textbf{58}, 675 (2008)
\bibitem{Toennies99}  R.E. Grisenti, W. Sch\"ollkopf, and J.P. Toennies, G. C. Hegerfeldt, and T. K\"ohler, \textit{Phys. Rev. Lett.} \textbf{83}, 1755 (1999)
\bibitem{Cronin05}  J.D. Perreault, and A.D. Cronin, \textit{Phys. Rev. Lett.} \textbf{95}, 133201 (2005)
\bibitem{Shimizu01} F. Shimizu, \textit{Phys. Rev. Lett.} \textbf{86}, 987 (2001)
\bibitem{Druzhinina03} V. Druzhinina and M. DeKieviet, \textit{Phys. Rev. Lett.} \textbf{91}, 193202 (2003)
\bibitem{Pasquini04} T.A. Pasquini, Y. Shin, C. Sanner, M. Saba, A. Schirotzek, D.E. Pritchard, and W. Ketterle, \textit{Phys. Rev. Lett.} \textbf{93}, 223201 (2004).
\bibitem{Schoellkopf08} B.S. Zhao, S.A. Schulz, S.A. Meek, G. Meijer, and W. Sch\"ollkopf, \textit{Phys. Rev. A} \textbf{78}, 010902(R) (2008).
\bibitem{Hinds92} V. Sandoghdar, C.I. Sukenik, E.A. Hinds, and S. Haroche \textit{Phys. Rev. Lett.} \textbf{68}, 3432 (1992)
\bibitem{Hinds93} C.I. Sukenik, M.G. Boshier, D. Cho, V. Sandoghdar, and E.A. Hinds, \textit{Phys. Rev. Lett.} \textbf{70}, 560 (1993)
\bibitem{Failache99} H. Failache, S. Saltiel, M. Fichet, D. Bloch, and M. Ducloy, \textit{Phys. Rev. Lett.} \textbf{83}, 5467 (1999)
\bibitem{Fichet07} M. Fichet et al., \textit{Eur. Phys. Lett.} \textbf{77}, 54001 (2007)
\bibitem{Aspect96} A. Landragin et al., \textit{Phys. Rev. Lett.} \textbf{77}, 1464 (1996)
\bibitem{Obrecht07} J.M. Obrecht, R.J. Wild, M. Antezza, L.P. Pitaevskii, S. Stringari, and E.A. Cornell, \textit{Phys. Rev. Lett.} \textbf{98}, 063201 (2007)
\bibitem{Grimm00} R. Grimm, M. Weidem\"uller, and Y.B. Ovchinnikov, \textit{Adv. At., Mol., Opt. Phys.} \textbf{42}, 95 (2000)
\bibitem{Bender09} H. Bender, P. Courteille, C. Zimmermann, and S. Slama, \textit{Appl. Phys. B} \textbf{96}, 275 (2009)
\bibitem{Kallush05}  S. Kallush, B. Segev, and R. C$\hat{\textrm{o}}$t\'e, \textit{Eur. Phys. J. D} \textbf{35}, 3 (2005).
\bibitem{Friedrich02}  H. Friedrich, G. Jacoby, and C.G. Meister, \textit{Phys. Rev. A} \textbf{65}, 032902 (2002)
\bibitem{Safronova04}  M.S. Safronova, Carl J. Williams, and Charles W. Clark, \textit{Phys. Rev. A} \textbf{69}, 022509 (2004).
\bibitem{Palik85}  E.D. Palik, \textit{Handbook of Optical Constants of Solids}, (Academic Press Inc., London, 1985).
\bibitem{footnote1}  Simpler fit
functions have also been tested. It has turned out that with the
given statistical errors of the data points e.g. a quadratic fit
function cannot be discerned from the general fit function, although
it does not fulfil condition (3).
\bibitem{McGuirk04}  J. M. McGuirk, D. M. Harber, J.M. Obrecht, and E.A. Cornell, \textit{Phys. Rev. A} \textbf{69}, 062905 (2004).
\bibitem{Wilde06}  A. Hatakeyama, M. Wilde, and K. Fukatani, \textit{e-J. Surf. Sci. Nanotech.} \textbf{4}, 63 (2006).
\bibitem{Efimov06}  A.M. Evimov, and V.G. Pogareva, \textit{Chem. Geo.} \textbf{229}, 198 (2006).


\end{thebibliography}
\end{document}